# Slip Length Measurement in Rectangular Graphene Nanochannels with a 3D Flow Analysis

Kuan-Ting Chen[1], Qin-Yi Li[1,4,*], Takeshi Omori[2], Yasutaka Yamaguchi[3,5], Tatsuya Ikuta[1,4], and Koji Takahashi[1,4,*]

[1] Department of Aeronautics and Astronautics, Kyushu University, Fukuoka 819-0395, Japan
[2] Department of Mechanical Engineering, Osaka City University, 3-3-138 Sugimoto, Sumiyoshi, Osaka, Osaka 558-8585, Japan
[3] Department of Mechanical Engineering, Osaka University, 2-1 Yamadaoka, Suita, Osaka 565-0871, Japan
[4] International Institute for Carbon-Neutral Energy Research (WPI-I2CNER), Kyushu University, Japan
[5] Water Frontier Research Center (WaTUS), Research Institute for Science & Technology, Tokyo University of Science, 1-3 Kagurazaka, Shinjuku-ku, Tokyo 162-8601, Japan
[*] Corresponding authors, Qin-Yi Li: qinyi.li@aero.kyushu-u.ac.jp

Koji Takahashi: takahashi@aero.kyushu-u.ac.jp

**ABSTRACT:**

Although many molecular dynamics simulations have been conducted on slip flow on graphene, experimental efforts remain very limited and our understanding of the flow friction on graphene remains far from sufficient. Here, to accurately measure the slip length in rectangular nanochannels, we develop a 3D capillary flow model that fully considers the nonuniform cross-section velocity profile, slip boundary conditions, and the dynamic contact angle. We show that the 3D analysis is necessary even for a channel with a width/height ratio of 100. We fabricated graphene nanochannels with 45-nm depth and 5-μm width, and measured slip lengths of about 30 to 40 nm using this 3D flow model. We also reevaluated the slip-length data for graphene obtained from capillary filling experiments in the literature: 30 nm instead of originally claimed 45 nm for a 25-nm-deep channel, and 47 nm instead of 60 nm for an 8.5-nm-deep channel. We discover a smaller slip length than existing experimental measurements due to our full 3D flow analysis considered in our method. This work presents a rigorous analysis approach while also providing a better understanding of slip flow in graphene nanochannels, which will benefit further innovation in nanofluidic applications, including electronics cooling and biomedical chips.

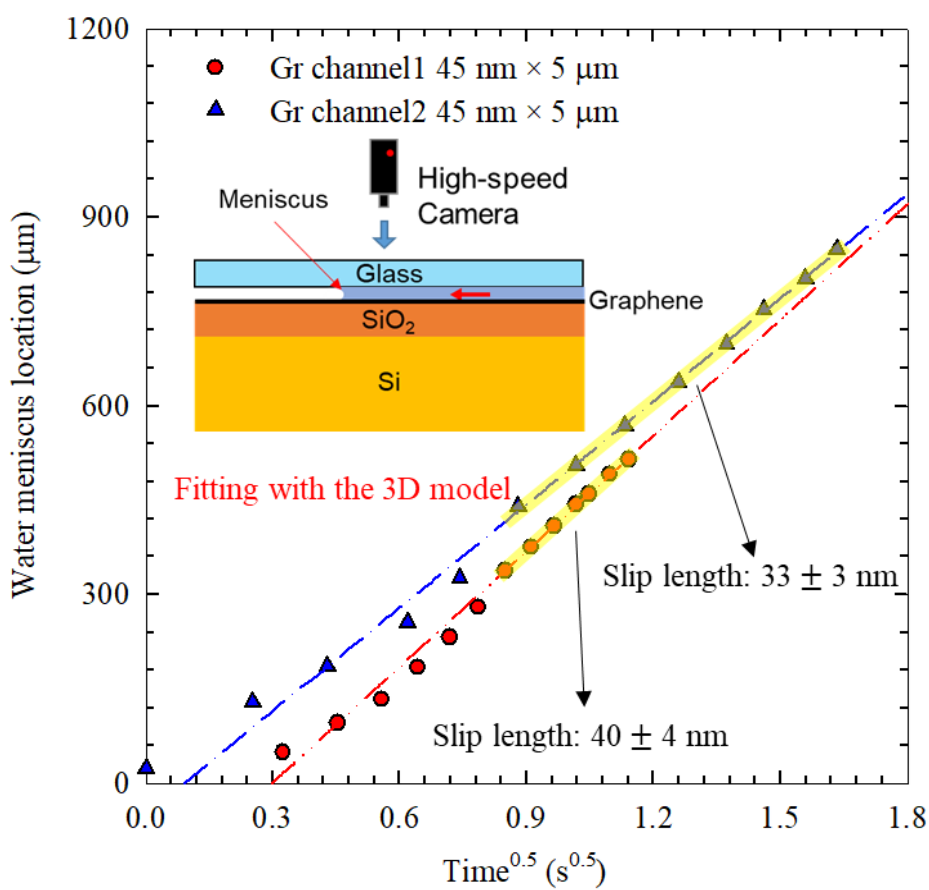

Table of Content

## 1. INTRODUCTION

Understanding the flow characteristics of fluid on the nanoscale has become critical for the optimization of nanoscale electronics cooling and biomedical devices [1]. Carbon nanotubes (CNTs) are popular devices for studying the fluid behaviors at the nanoscale [2-10]. Holt *et al.* [3] and Majumder *et al.* [4] studied the slip flow of water in 1.6-nm- and 7-nm-diameter CNT membranes, respectively. The observed flow rate was two to five orders faster than that predicted by the Hagen–Poiseuille relation [11], which assumes continuum governing equations with the nonslip boundary conditions. Since then, many researchers have focused on the molecular dynamics (MD) analysis of the slip properties of water on the nanoscale, specifically in CNTs. Thomas and McGaughey [5] predicted that the slip lengths in single-walled CNTs converge to 30 nm when increasing the diameter of the CNT to infinity [5], and increase up to 500 nm with decreasing diameter. Other MD simulation studies showed that the slip length in CNTs depends on the viscosity and the interaction between water and carbon [6-10]. Apart from CNTs, simulation research also focuses on graphene nanochannels, which may be more suitable for lab-on-chip applications. Kannam *et al.* [12] found a slip length of 60 nm in a planar graphene nanochannel using equilibrium MD simulations and pointed out that the nonequilibrium molecular dynamics (NEMD) simulations have large statistical errors when determining the curvature of the velocity profile. Wagemann *et al.* [13] investigated the dependence of slip length on the shear stress in a graphene nanochannel using NEMD simulations and found the slip length converged to 50 nm under low shear stress. However, another MD simulation reported a slip length of 3 nm in graphene nanochannels [14]. The large deviation in the MD predictions of the slip length on graphene has not been clarified yet.

Experimental research on slip flow in graphene nanochannels was absent until the work of Xie *et al.* in 2018 [15]. These authors developed a 2D flow model based on the Washburn theory [16-17] to determine the slip length of water on graphene in graphene/$SiO_2$ hybrid rectangular nanochannels with a height range from 25 to 105 nm and a width of 5 µm. The slip length was determined by comparing water capillary filling in the graphene and $SiO_2$ sections of the hybrid nanochannel. The obtained slip length ranged from 0.5 to 200 nm and was independent of the height of the nanochannels. The physical reasons for the widely scattered slip-length data have not yet been clarified. Notably, Xie *et al.* assumed the same pressure drop in both the $SiO_2$ and graphene sections of the hybrid nanochannels, which can cause large uncertainty. Very recently, Keerthi *et al.* [18] investigated slip flow in graphite nanochannels with a width of 130 nm and height range of 0.68–8.5 nm. Because of the difficulties in observing the meniscus with such a small channel height, microgravimetric analysis was used to determine the rate of water loss in the water reservoir. This was considered to be the flow rate inside the nanochannel because the water flows through the channel and evaporates at the outlet of the nanochannel resulting in mass loss. Using the velocity profile obtained from the Navier–Stokes equation for the 2D analysis and modified with the slip term, they obtained the relation between the mass flux and filling distance. Comparison of the 2D flow model and observed mass loss rate led to a calculated slip length of ~60 nm when the depth was between 2.38 and 8.5 nm.

These rectangular nanochannels fabricated with micro-electro-mechanical-systems (MEMS) processes have micrometer-scale widths that are usually two orders of magnitude larger than their depths. To the best of our knowledge, all the experimental studies on the flow in rectangular nanochannels assumed a 2D Washburn model that neglected the velocity distribution in the width direction, due to the inertial thinking on this kind of nanochannel with a high aspect ratio; however, this assumption is questionable. As for some flow experiments in nanopores or $SiO_2$ nanochannels with a large width/depth ratio, the observed flow rates followed the 2D Washburn model qualitatively but were all lower than the predicted values [19-25]. For non-Newtonian liquids, Cao *et al.* [26] in 2016 modified the Washburn model with the polymer rheological model, which agreed with the rise of the meniscus in nanopores. For Newtonian liquids, some researchers have investigated potential reasons for this difference, including the electroviscous effect [19-22], channel deformation [23-24], and the effect of the dynamic contact angle [25]. However, all the effects mentioned here are either insufficient or unsuitable to explain the lower flow rate than that predicted by the 2D model. It is necessary to develop a reliable analysis method to accurately measure the slip length in rectangular nanochannels.

In the current work, we develop a rigorous 3D model to measure the slip length that considers slip boundary conditions, a nonuniform velocity profile in both the width and depth directions, and the dynamic contact angle. We show that a nonuniform cross-section velocity profile has a significant effect on the flow rate even in a nanochannel with a width-to-depth ratio of 100. Our model applies to nanochannels with different slip velocities and contact angles on different walls. Graphene nanochannels with 45-nm depth and 5-μm width were fabricated, and the slip lengths on graphene were measured by comparing the newly developed 3D flow model with the observed flow rates.

## 2. EXPERIMENTAL

The graphene nanochannels were fabricated on a $SiO_2$/Si wafer using electron-beam lithography (EBL) and wet etching, followed by graphene transfer from the copper substrate and anodic bonding of the glass cover, as illustrated in Figure 1a. First, the silicon wafer with a thermally grown $SiO_2$ layer, a sacrificing Cr layer, and an electron beam resistor (EBR) layer went through the EBL process for the patterning of the nanochannels. The sacrificing Cr layer was etched by the $Fe(NO_3)_3$ solution, and part of the $SiO_2$ layer was etched by buffered hydrofluoric acid (BHF). Next, single-layer graphene (bought from Graphene Platform Corp., Japan) was transferred on the nanochannels with a wet transfer method.[27]. Finally, a Pyrex glass cover was sealed on the top of the nanochannels by an anodic bonding process.

The test section (a graphene nanochannel bonded with Pyrex glass on the top) was set under a microscope with a high-speed camera (FASTCAM Mini AX50, Photron) installed to visualize the capillary filling of water and to track the position of the meniscus, as shown in Figure 1b to Figure 1d. The high-speed camera began recording, then a water droplet was dropped on the water reservoir but kept a distance from the entrance of the nanochannel to prevent the effect of the gravity of the dropping droplet. The droplet will eventually contact the entrance as it slowly spreads on the surface and the filling process of the water started as a result of the capillary force.

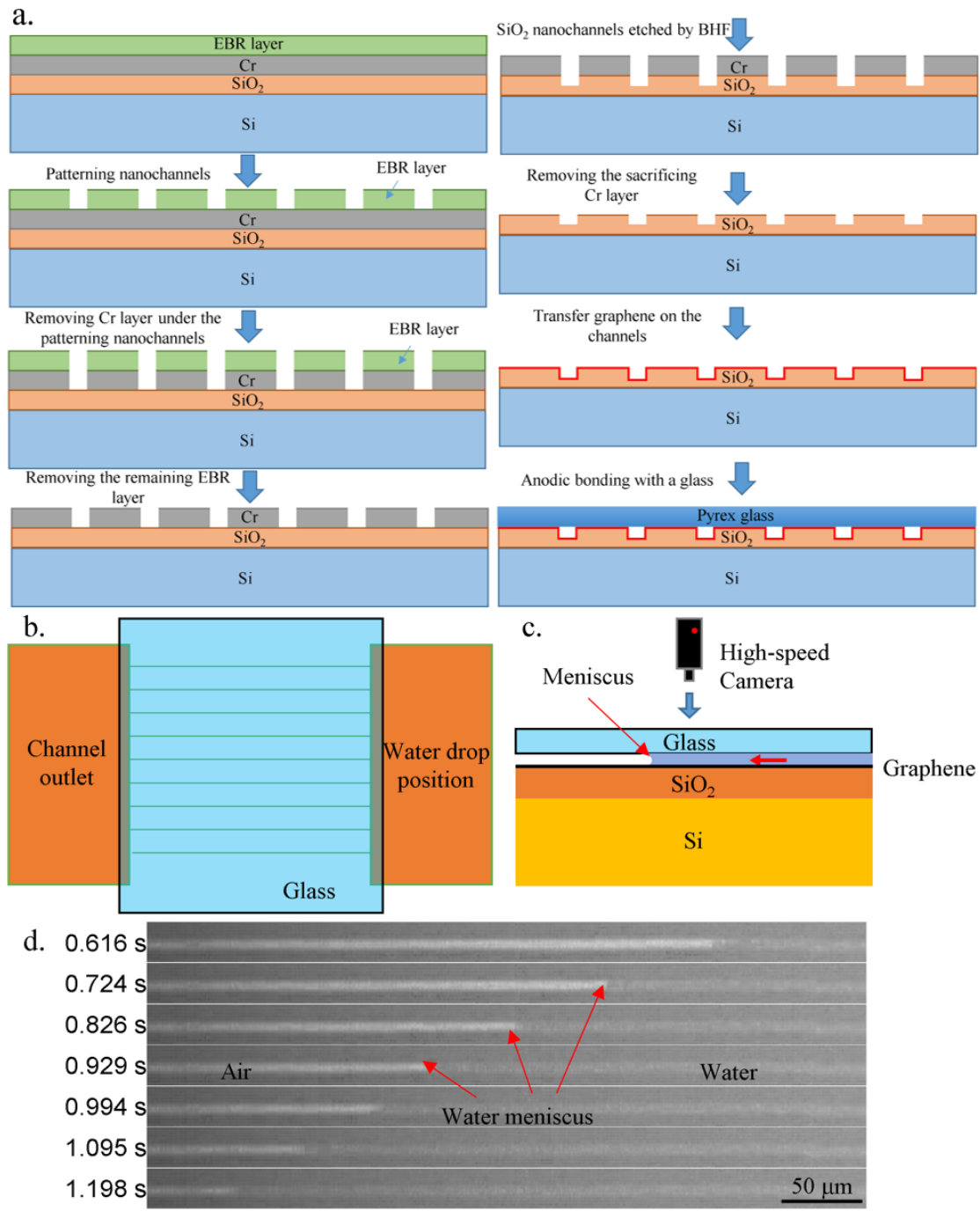

**Figure 1.** The experimental system for observing meniscus movement: (a) Fabrication process of the graphene nanochannels. (b) top view, (c) side view, and (d) high-speed-camera image of capillary filling in a 45-nm-deep and 5-µm-wide graphene nanochannel (Gr channel1).

The graphene nanochannel on the test section has a width of 5 µm, as shown in Figure 2b and Figure 2d. The nanochannel length was 2 mm with a depth of 45.0 ± 0.5 nm. The surface roughness was measured in five different spots with an area of 1 µm$^2$, and the average roughness values were 0.33 nm for the SiO$_2$ nanochannel before graphene transfer, and 0.89 nm for the graphene nanochannel. Higher surface roughness of graphene than SiO$_2$ was also observed by Xie *et al.* [15], who measured 0.47 nm for SiO$_2$ and 0.98 nm for graphene. The increased surface roughness on graphene surfaces invariably occurred during the transfer process. The rougher graphene surface was also visible in 3D images, as observed in Figure 2a and Figure 2b. At first glance of the geometry as seen in the profile analysis in Figure 2d, the occurrence of slip on such a rough surface might not seem possible. However, by dividing the width by the number of the peaks in the height profile in Figure 2d, we notice a flat or curved area of about 200 nm between every two maximum heights, indicating that such a graphene surface is sufficiently flat for slip flow of water to occur.

The graphene surface has a low level of defects, as evident from the negligible D peak in the Raman spectrum in Figure 3a [28-29]. The graphene nanochannel was fabricated by transferring graphene on 5-µm-width SiO$_2$ nanochannels using the wet transfer method [27]. Another chip that went through the same etching process to simulate the surface condition inside nanochannels was used to measure the static contact angle of water. The water contact angle on Pyrex glass was also measured because the top of the channel was covered by the glass. The water contact angles on Pyrex glass and graphene on SiO$_2$ were measured to be 68.0° and 83.7°, respectively, as shown in Figure 3b and Figure 3c. The static contact angle on graphene was measured after 2 days of the graphene transfer because the contact angle on graphene varies during the first 24 hours of exposure to the atmosphere [18, 30]. In the literature, the water contact angle on graphene was measured in the range of 73°–84° after exposure to air for 1 day by Kozbial *et al.* [30] and in the range of 62°–85° after 4 days of exposure to air by Keerthi *et al.* [18].

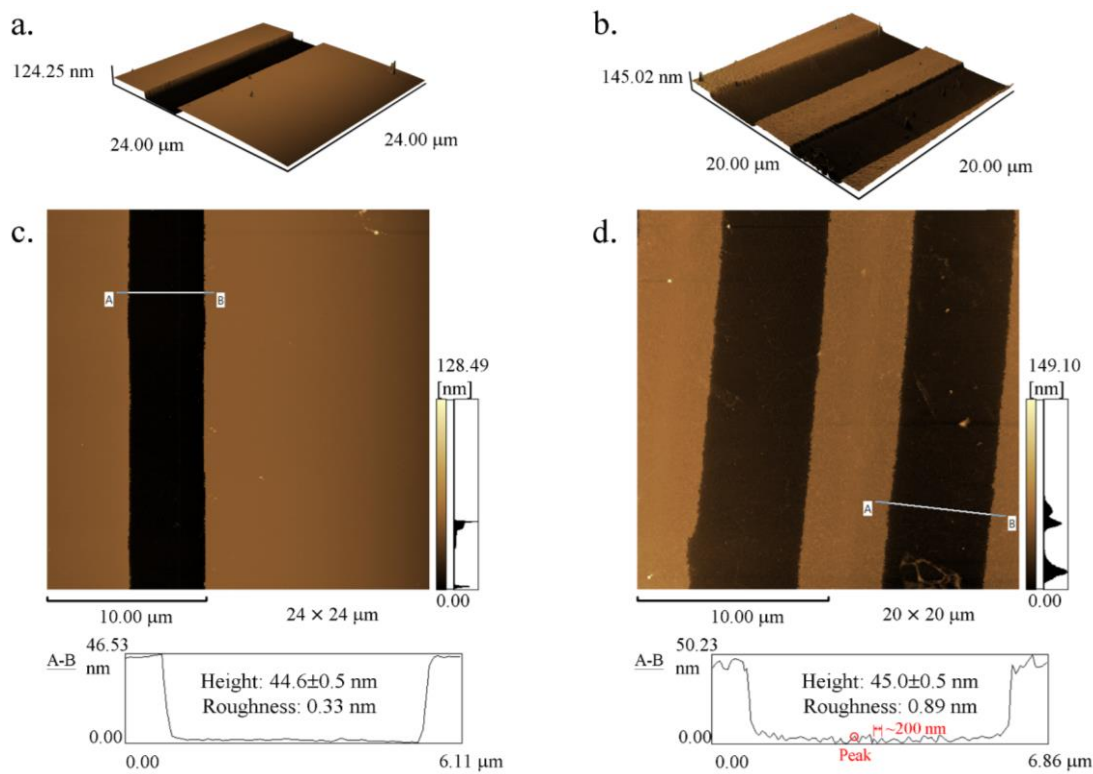

**Figure 2.** Atomic force microscopy (AFM) scanning results of different nanochannels. (a) 3D image of a 5-µm-wide SiO$_2$ nanochannel. (b) 3D image of 5-µm-wide graphene nanochannels. (c) Top and cross-sectional view of 5-µm-wide SiO$_2$ nanochannel. (d) Top and cross-sectional view of the 5-µm-wide graphene nanochannels.

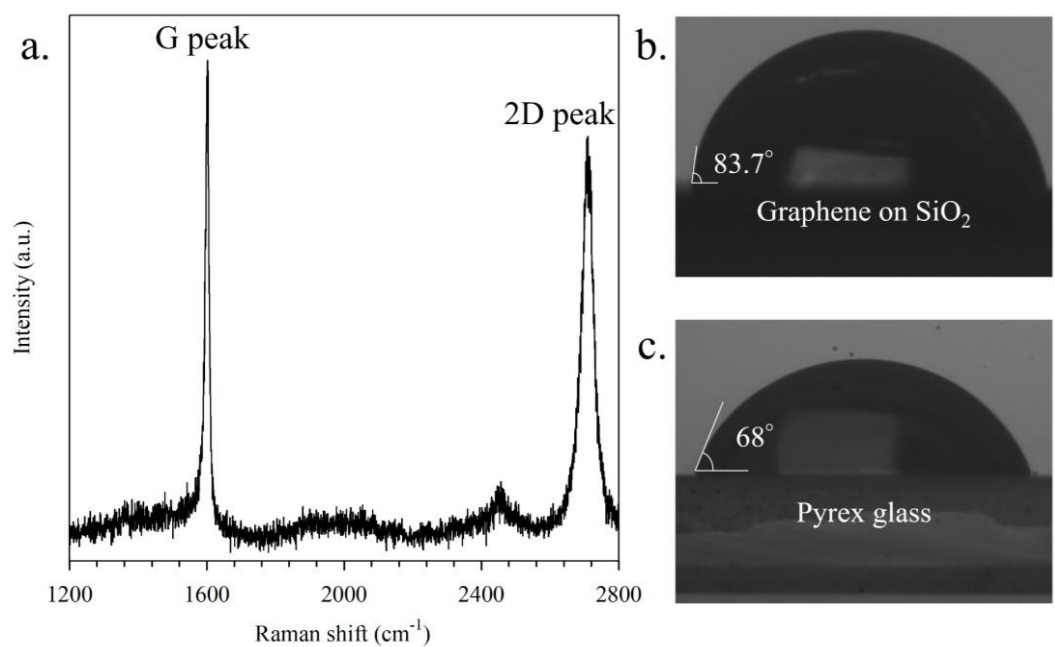

**Figure 3.** (a) Raman Spectrum of the present graphene surface. (b, c) The static contact angles of water on graphene and Pyrex glass.

# 3. THEORY

For capillary filling of a liquid in a rectangular channel, the meniscus movement is governed by the balance between the capillary force due to the surface tension and the viscous force, as shown in Figure 4a. To determine the meniscus location over time, it is necessary to know the physical properties of the liquid and the boundary conditions on the channel wall (*e.g.*, nonslip or slip boundary condition). The commonly used Washburn equation [16] for predicting the water meniscus movement in micro/nanochannels assumes a uniform velocity profile across the channel-width direction; in contrast, we developed a model considering a nonuniform velocity profile in the directions of both the channel width and depth.

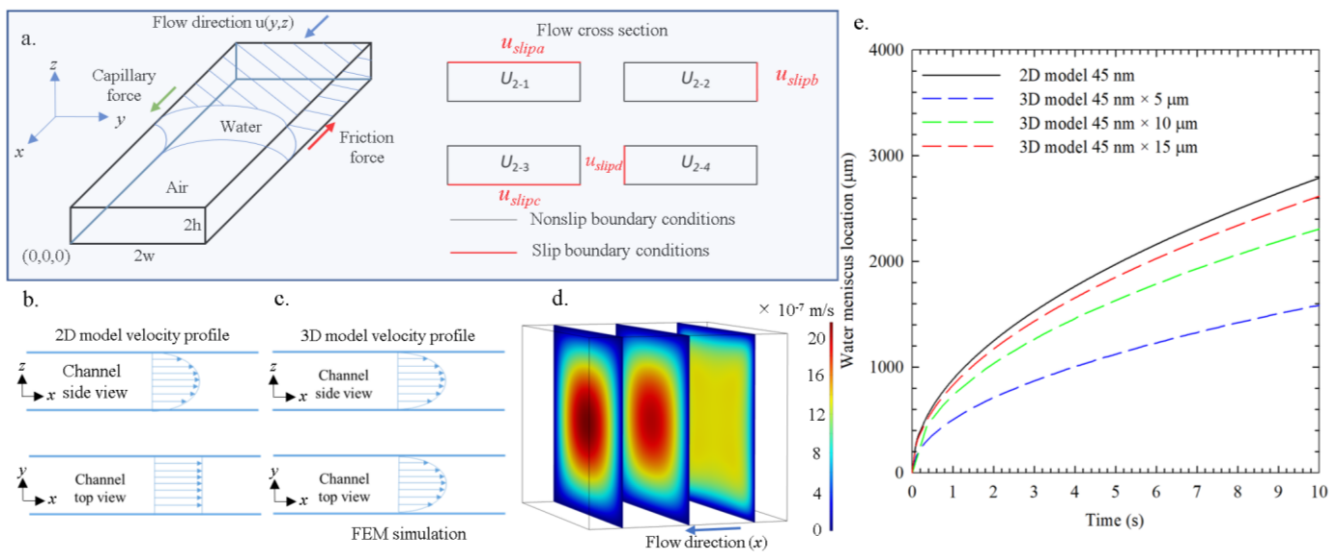

**Figure 4.** Capillary filling in a nanochannel. (a) Coordinates and decomposition of slip boundary conditions. (b) The assumption for the velocity profile in the 2D Washburn model. (c) Schematic for the velocity profile in the 3D flow model. (d) 3D velocity profiles obtained from the finite-element-method (FEM) simulations of a 50-nm-wide, 45-nm-deep channel. (e) Water meniscus locations versus the flow time calculated by the analytical solutions to the 2D and 3D models for different channel geometries (45 nm deep, 5-15 μm wide).

The traditional model used in the literature only considers the two-dimensional velocity profile as shown in Figure 4b and assumes that the velocity profile is uniform across the width direction of the channel, as shown in Figure 4b. Derived from the Navier–Stokes equation with the nonslip boundary conditions for a developed laminar flow with 2D analysis, the following velocity profile is obtained as shown in Eq. (1):

$$u = \frac{1}{2\mu}\left(\frac{\partial p}{\partial x}\right)(z^2 - h^2) \tag{1}$$

where $u$ is the velocity profile, $\mu$ is the viscosity, $p$ is the capillary pressure, $x$ is the fluid-flow direction, and $h$ is half of the channel height.

By defining the meniscus velocity to be the average flow velocity, and using the Laplace-young equation for the pressure drop on the meniscus, one derives the Washburn equation as shown in Eq. (2), also known as the modified Washburn equation [31] used for a rectangular channel:

$$l = \sqrt{\frac{2ht\sigma \cos\theta}{3\mu}} \tag{2}$$

where $l$ is the meniscus location from the inlet; $t$ is time; and $\sigma$ and $\theta$ are the surface tension and contact angle of the liquid, respectively (see the detailed derivation of the 2D model in Appendix A.1).

Although the meniscus movement can be determined using Eq. (2), the equation often overpredicts the results in nanochannels. The assumption of a uniform velocity profile along the y-direction across the channel cannot be justified in general. Therefore, we developed a 3D model that considers non-uniform velocity profiles both along the y- and z-directions across the channel, as shown in Figure 4c. Considering the fully developed laminar flow with different slip velocities on different walls, and neglecting the effect of gravity, as shown in Figure 4a, the velocity profile can be obtained as shown in Eqs. (3)–(8) from the Navier–Stokes equation (see the detailed derivation in Appendix A.2).

$$u = u_1 + u_{2-1} + u_{2-2} + u_{2-3} + u_{2-4} \tag{3}$$

$$u_1 = \sum_{n=0}^{\infty} \frac{16\Delta p (-1)^n h^2}{\mu l [\pi(2n+1)]^3} \left(1 - \frac{\cosh\left[\frac{(2n+1)\pi(y-w)}{2h}\right]}{\cosh\left[\frac{(2n+1)\pi w}{2h}\right]}\right) \cos\left[\frac{(2n+1)(z-h)\pi}{2h}\right] \tag{4}$$

$$u_{2-1} = \sum_{n=0}^{\infty} \frac{u_{slipa}}{w} \cdot \frac{1}{\sin\left(\frac{n\pi h}{w}\right)} \cdot \left[\frac{2w}{n\pi}\cdot \cosh(n\pi) - \frac{2w}{n\pi}\right] \cdot \sinh\left(\frac{n\pi y}{2w}\right) \cdot \sin\left(\frac{n\pi z}{2w}\right) \tag{5}$$

$$u_{2-2} = \sum_{n=0}^{\infty} \frac{u_{slipb}}{h} \cdot \frac{1}{\sinh\left(\frac{n\pi w}{h}\right)} \cdot \left[-\frac{2h}{n\pi}\cdot \cos(n\pi) + \frac{2h}{n\pi}\right] \cdot \sinh\left(\frac{n\pi y}{2h}\right) \cdot \sin\left(\frac{n\pi z}{2h}\right) \tag{6}$$

$$u_{2-3} = \sum_{n=0}^{\infty} \frac{u_{slipc}}{w} \cdot \frac{1}{\sin\left(-\frac{n\pi h}{w}\right)} \cdot \left[\frac{2w}{n\pi}\cdot \cosh(n\pi) - \frac{2w}{n\pi}\right] \cdot \sinh\left(\frac{n\pi y}{2w}\right) \cdot \sin\left(\frac{n\pi(-z+2h)}{2w}\right) \tag{7}$$

$$u_{2-4} = \sum_{n=0}^{\infty} \frac{u_{slipd}}{h} \cdot \frac{1}{\sinh\left(-\frac{n\pi w}{h}\right)} \cdot \left[-\frac{2h}{n\pi}\cdot \cos(n\pi) + \frac{2h}{n\pi}\right] \cdot \sinh\left(\frac{n\pi(-y+2w)}{2h}\right) \cdot \sin\left(\frac{n\pi z}{2h}\right) \tag{8}$$

Here, $u_{slipa}$, $u_{slipb}$, $u_{slipc}$, and $u_{slipd}$ are different slip velocities on different walls, as shown in Figure 4a.

The slip length, $L_{slip}$, is related to the slip velocity by Eq. (9):

$$\frac{u_{slip}}{L_{slip}} = \left.\frac{\partial u}{\partial z}\right|_{z=0, z=2h} \tag{9}$$

After obtaining the velocity profiles using Eqs. (4)–(8), the relation between the moving meniscus location and time for a nanochannel with the wall having slip boundary conditions can be obtained (Appendix Eqs. (A-27)–(A-32)) by the integration of the velocity profile. For a nanochannel with all the channel walls having different slip boundary conditions, the water flow movement in the nanochannels can be expressed using Eq. (10):

$$l = \sqrt{\left[\frac{8\cdot \Delta p \cdot h^2}{\mu w \pi^3} \cdot \sum_{n=0}^{\infty} \frac{(-1)^n}{(2n+1)^3}\left(2w - \frac{4h}{(2n+1)\pi}\cdot \tanh\frac{(2n+1)\cdot \pi w}{2h}\right) \cdot \left[\frac{4h}{(2n+1)\cdot \pi}\cdot \sin\frac{(2n+1)\cdot \pi}{2}\right] + D_2 + D_3 + D_4 + D_5\right]\cdot t} \tag{10}$$

where $D_2$, $D_3$, $D_4$, and $D_5$ represent the slip term (Appendix Eqs. (A-29)–(A-32)). Note the linear relation between $l$ and $t^{0.5}$ in a fully developed laminar flow in Eq. (10). However, the exact region needed to form the fully developed flow is unknown. According to Stange et al. [32], the inertial effects and the convective losses at the entrance dominate the early stage of capillary filling, resulting in a linear relationship between $l$ and $t^2$ or $l$ and $t$, rather than $l^2$ and $t$. In such a case, an entrance length ($l_0$) before the flow becomes fully developed is needed. The corresponding equation then becomes Eq. (11) when considering the slip boundary conditions on the wall. The fully developed flow starts when the time becomes 0.

$$l = \sqrt{\left[\frac{8 \cdot \Delta p \cdot h^2}{\mu w \pi^3} \cdot \sum_{n=0}^{\infty} \frac{(-1)^n}{(2n+1)^3}\left(2w - \frac{4h}{(2n+1)\pi} \cdot \tanh\frac{(2n+1)\cdot \pi w}{2h}\right) \cdot \left[\frac{4h}{(2n+1)\cdot \pi} \cdot \sin\frac{(2n+1)\cdot \pi}{2}\right] + D_2 + D_3 + D_4 + D_5\right] \cdot t} + l_0 \quad (11)$$

where $\Delta p$ is the capillary pressure under the assumption of different contact angles ($\theta_1$, $\theta_2$, $\theta_3$, and $\theta_4$) on different walls as shown in Eq. (12):

$$\Delta p = \frac{\sigma(2w \cdot \cos\theta_1 + 2w \cdot \cos\theta_2 + 2h \cdot \cos\theta_3 + 2h \cdot \cos\theta_4)}{4wh} \quad (12)$$

After obtaining the relationship between water meniscus location and meniscus flowing time, the slip length can be measured by the capillary flow experiment in the rectangular nanochannel. In our experiment, we recorded the meniscus movement in the nanochannel with time. In the fully developed flow region, the meniscus location ($l$) versus the square root of time ($t^{0.5}$) has a linear relationship as indicated by Eq. (11). We plot $l$ versus $t^{0.5}$ and fit the experimental data in the fully developed flow region for the slope of $dl/dt^{0.5}$, as shown in Fig. 5. This $dl/dt^{0.5}$ slope is related to the slip length and contact angles. With the contact angles on both the graphene and glass walls measured beforehand, we can extract the slip length from the slope of $dl/dt^{0.5}$ using Eq. (11). Note that the observation time of the meniscus movement should be long enough so that the $l$ vs. $t^{0.5}$ relation can become linear, which indicates that the flow becomes fully developed. Only the data points within the fully developed region should be used for the linear fitting, and in this way, the entrance effects in the initial stage can be easily eliminated.

The top side of the channel fabricated in this work was covered by glass, for which the contact angle was different from the graphene-coated walls. When all contact angles on all channel walls are 0, a much simpler equation is obtained as Appendix Eq. (A-36). The above analytical solution to the 3D flow model was verified by comparing it with the FEM simulation results, the details of which are provided in Supplementary Note S1.

The airflow pressure drop was not included in the model because the channel depth was smaller than the mean free path of the air. The mean free path of the air was calculated to range from 63.9 to 68 nm between 15.15 °C and 25.15 °C and at a relative humidity between 0% and 100% [33]. Because the nanochannels discussed in this work were less than 50 nm in depth, the air remaining in such a channel geometry can be considered as a rarefied gas, and the drop in its pressure can be considered negligible.

The effect of the dynamic contact angle can be seen using Eq. (A-34) in the Appendix. Because the velocity of water meniscus movement in such a small nanochannel is on the scale of hundreds of micrometers per second, the capillary number is extremely small and there is only a small change in contact angle. The dynamic contact angle change was calculated to be less than 0.2°, which affects the prediction of the meniscus movement by less than 2% and can be considered negligible.

Although a similar 3D model was previously developed and used to analyze the flow in a microchannel with a small width/depth ratio [34], our model is more detailed by considering different slip boundary conditions and contact angles on different walls, as well as the effects of the rarefied gas and the dynamic contact angle.

Figure 4e demonstrates the 3D effect on the capillary flow speed in rectangular nanochannels by showing the meniscus location through time for different channel sizes. The black solid line gives the meniscus movement calculated by the 2D flow model for water's capillary filling in a 45-nm-deep channel, which represents a channel with an infinite width-depth ratio. The red, green, and blue dash lines show the meniscus movement in nanochannels with a width-depth ratio of around 300, 200, and 100, respectively. The average flow velocities in the first ten seconds of the capillary filling calculated by the 3D model are 40%, 18%, and 7% lower than those calculated by the 2D model for the nanochannels with width-depth ratios of 100, 200, and 300, respectively. When the width-depth ratio is larger than 350, the deviation between the predicted flow velocities by the 2D and 3D models can be within 5%. For a width-depth ratio of 1000 and 10000, the deviations between the 2D and 3D models are as low as 0.7% and 0.002%, respectively. In summary, the 3D effect for capillary filling in rectangular nanochannels is important even when the width is two orders of magnitude larger than the depth, but negligible when the width-depth ratio is larger than 1000.

## 4. RESULTS AND DISCUSSION

The temporal changes in the location of the water meniscus in two graphene nanochannels with the same geometries, denoted as Gr channel1 and Gr channel2, were recorded with a high-speed camera, as shown in Figure 1c and Supplementary Fig. S2. The meniscus locations in both channels were plotted versus the square root of time in Figure 5a. As mentioned in Section 3, the initial effects and the convective losses at the entrance dominate the early stage of capillary filling [32]. The flowing distance and the square root of time will have a linear relationship when the time is long enough, and the flow becomes fully developed. The $dl$–$dt^{1/2}$ slope for Gr channel1 and Gr channel2 in the fully developed region are $613 \pm 7$ μm/s$^{0.5}$ and $556 \pm 7$ μm/s$^{0.5}$, respectively. Using the static contact angles measured for graphene and Pyrex glass (assumed nonslip boundary conditions) shown in Figure 3b and Figure 3c, and calculated with Eq. (9) and Eq. (10), the slip lengths in both 45-nm-deep graphene nanochannels were calculated to be $40 \pm 4$ nm and $33 \pm 3$ nm.

Next, we used the 3D model to reevaluate the slip length data obtained by the 2D model, available in the literature. Xie *et al.* [15] compared the capillary flow in SiO$_2$ and graphene nanochannels with a 2D model. However, they did not measure the contact angles, but assumed the same contact angle for both SiO$_2$ and graphene nanochannels and canceled the capillary pressure term in the calculation. Accordingly, they calculated the slip length in a 25-nm-deep graphene nanochannel to be $45 \pm 2$ nm. Here, we fit the fully developed region of their data [15] with our 3D model by assuming the same contact angles for graphene and glass as measured in our work (see the fitting curve in Supplementary Fig. S1). The linear fitting slope is

463 ± 13 µm/s$^{0.5}$, resulting in a slip length of 30 ± 5 nm. Therefore, by neglecting the 3D effect, the slip length acquired in the work of Xie *et al.* [15] is 50% higher than our current work.

In the work by Keerthi and co-workers [18], the meniscus position was fixed in the nanochannel. They considered the water loss rate as the mass flow rate inside their nanochannel and used a 2D model to analyze the slip length as mentioned in the introduction. We used 3D analysis to modify the mass flux model used in their study to recalculate the slip length in the graphite nanochannel with a 130-nm width and 8.5-nm depth (*i.e.*, the same dimensions used in their study) in a mass flow rate of 4.3×10$^{-11}$ g/s. By changing the velocity profile into 3D analysis results, as described by Eqs. (4)–(8), and with the integration of a velocity profile, as described by Eq. (A-29), a 3D expression for the mass flow rate can be obtained to recalculate the slip length. With this 3D model of mass flux and the same graphene contact angle measured as in their research (85°), a slip length of 47 ± 16 nm was obtained. Therefore, the slip length in the work of Keerthi *et al.* [18] is 28% higher than our current work.

In Figure 5b, the hollow columns of "Gr channel1" and "Gr channel2" represent the measured slip lengths extracted with the 3D flow model in our present work. The solid columns in "Xie, 2018" and "Keerthi, 2021" represent the only two literature works on the slip length in graphene nanochannels. As explained in the introduction part, these two experiments extracted the slip length with a 2D flow model, which caused large errors. We re-analyzed the original flow data in these two references with our 3D model, and the strip-filled columns in "Xie, 2018" and "Keerthi, 2021" represent the corrected slip lengths using our 3D model. The uncertainty analysis is provided in Supplementary Note S2. Our measured slip lengths of 40 ± 4 nm and 33 ± 3 nm are close to those corrected by the 3D flow model using the data from the literature, which approximately fall in the range of 30 to 50 nm and agree with the MD simulation results of the slip lengths for infinitely large CNTs [5] and graphene [13]. The 2D flow model significantly overestimated the slip length in graphene nanochannels, by 50% for the work of Xie *et al.* [15] and 28% for the work of Keerthi *et al.* [18]. The deviations between the slip length data can be attributed to several reasons. One reason is the surface roughness, because simulations showed that a smooth surface and flattened free energy landscape of the water contact layer are important to lowering the friction [35-37]. The surface charge density is another factor that affects the water slip length on graphene; as proposed in previous simulation research [15, 38], the slip length decreases as the surface charge increases because of a higher coverage of functional groups on the graphene surface. Another important reason is the variations in the quality of the graphene surface. Even when using the same wet transferring method [27] and the same graphene source to transfer graphene to the $SiO_2$ substrate, it is impossible to fabricate the same graphene surface on the same substrate; some cracking or gathering of graphene is inevitable [15], especially for deeper channels; this is different from the perfectly uniform graphene surface assumed in previous simulation research [12-14].

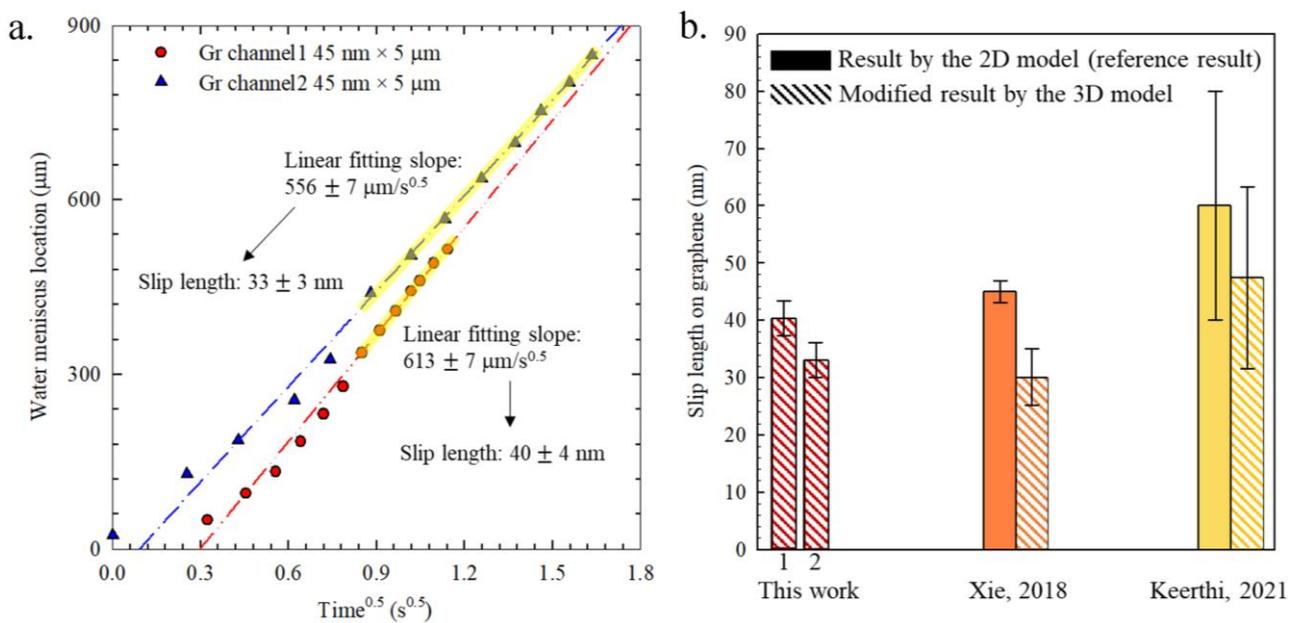

**Figure 5.** (a) Water meniscus location changing with time and the linear fitting of the fully developed region of our 45-nm-deep graphene nanochannels. (b) Slip lengths in our work with the 3D model and in the references with both 2D and 3D models.

## 5. CONCLUSION

In this work, we measured the slip length in rectangular graphene nanochannels from capillary flow analysis with a 3D flow analysis, which considers different slip boundary conditions and contact angles on different walls, effect of the rarefied gas, and the effect of dynamic contact angles. We demonstrated that the 3D effect in the flow model cannot be neglected for the extraction of the slip length even if the nanochannel width is two orders of magnitude larger than the depth. We fabricated graphene nanochannels with 45-nm depth and 5-µm width, and recorded temporal changes of the water meniscus location in two graphene nanochannels. We measured slip lengths of $40 \pm 4$ nm and $33 \pm 3$ nm by comparing the capillary flow rate with the 3D flow model. Moreover, the slip lengths for graphene reported in previous research that neglected the 3D effect were reevaluated using our model, from $45 \pm 2$ nm to $30 \pm 5$ nm and from $60 \pm 20$ nm to $47 \pm 16$ nm in 25-nm-deep and 8.5-nm-deep graphene nanochannels, respectively. These experimental data of the water slip length on graphene approximately fall in the range of 30 to 50 nm, in agreement with some important MD simulation results. The variation in the measured slip lengths may originate from the different surface roughness, surface charge density, and graphene quality, which requires further investigation. Our work provides not only a rigorous analysis approach for the slip length measurement in nanochannels, but also a better understanding of water slip flow on graphene surfaces, which will benefit future innovation of nanofluidic applications.

**APPENDIX**

**A.1 Derivation of 2D Washburn model used in the literature**

Consider a fluid flow inside a square channel with channel width $2w$ and channel depth $2h$ with four walls with the nonslip boundary conditions, as shown in Figure 4b. The Navier–Stokes equation in such a square channel can be expressed by Eq. (A-1):

$$\mu\left(\frac{\partial^2 u}{\partial y^2} + \frac{\partial^2 u}{\partial z^2} + \frac{\partial^2 u}{\partial x^2}\right) + \rho g_x - \frac{\partial p}{\partial x} = \rho\left(\frac{\partial u}{\partial t} + u\frac{\partial u}{\partial x} + v\frac{\partial u}{\partial y} + w\frac{\partial u}{\partial z}\right) \quad \text{(A-1)}$$

where $\propto$ is the viscosity; $u$, $v$, and $w$ represent the velocity components in the $x$-, $y$-, and $z$-directions, respectively; $\rho$ is the density of the fluid; $g$ is the gravitational acceleration; and $p$ is a capillary-pressure term. The assumption of no velocity in the $y$- or $z$-directions makes the $v$ and $w$ values equal 0. Assuming a uniform velocity profile across the $y$-direction, and neglecting the gravity effect, the Navier–Stokes equation can be simplified as Eq. (A-2).

$$\mu\left(\frac{\partial^2 u}{\partial z^2}\right) - \frac{\partial p}{\partial x} = 0 \qquad (A-2)$$

The nonslip boundary conditions shown in Eq. (A-3), where $h$ is half of the channel height, were used in this 2D model; the velocity profile as shown in Eq. (1) in the main text can be obtained using these boundary conditions:

$$z = -h, z = h : u = 0 \qquad (A-3)$$

The volume flow rate $Q$, which passes through two parallel plates, can be obtained by integrating the velocity distribution as shown in Eq. (A-4):

$$Q = \int_{-h}^{h}\int_{-w}^{w} u\, dy\, dz = \int_{-w}^{w}\int_{-h}^{h} \frac{1}{2\mu}\left(\frac{\partial p}{\partial x}\right)(z^2 - h^2)\, dz\, dy = \int_{-w}^{w} -\frac{2h^3}{3\mu}\left(\frac{\partial p}{\partial x}\right) dy = -\frac{4wh^3}{3\mu}\left(\frac{\partial p}{\partial x}\right) \qquad (A-4)$$

The pressure gradient is negative because the pressure decreases along the direction of the flow. It can also be expressed as Eq. (A-5), assuming the pressure gradient to be the pressure drop between two points at a distance $l$ apart. The flow rate can then be written as shown in Eq. (A-6).

$$-\left(\frac{\partial p}{\partial x}\right) = \frac{\Delta p}{l} \qquad (A-5)$$

$$Q = \frac{4wh^3}{3\mu}\left(\frac{\Delta p}{l}\right) \qquad (A-6)$$

The flow rate can also be expressed as the cross-sectional area times the length moved per unit time, as written in Eq. (A-7). After integration, Eq. (A-8) is obtained.

$$Q = \frac{dV}{dt} = \frac{4wh^3}{3\mu}\left(\frac{\Delta p}{l}\right) = 2w \cdot 2h \cdot \frac{dl}{dt} \qquad (A-7)$$

$$l = \sqrt{\frac{2h^2 t}{3\mu}\Delta p} \qquad (A-8)$$

The capillary force is the dominant force in the pressure term, which can be expressed as Eq. (A-9) in a square channel, as shown in Figure 4a; here, $\sigma$ and $\theta$ are the surface tension and contact angle of the liquid, respectively. After substituting the pressure term in Eq. (A-8), the relationship between time and the meniscus location can be obtained, as shown in Eq. (2) in the main text, also known as the modified Washburn equation [31] used for the rectangular channel.

$$\Delta p = \frac{\sigma \cos \theta}{h} \tag{A-9}$$

**A.2 Derivation of the 3D flow model**

With the assumption of no velocity gradient in the *y*- or *z*-directions, the *v* and *w* values are 0. In addition, the flow could be considered as fully developed laminar flow because the meniscus flow movement used in previous studies was extremely slow [21-25, 39] (hundreds of micrometers per second), which results in a much low flow rate and a calculated entrance length that is four orders of magnitude smaller than 1 nm. In such a case, the flow could be considered as a fully developed laminar flow, and it is safe to consider that at time zero, the capillary filling length is zero. Finally, by neglecting the gravity effect because it has an extremely small *Re*, the Navier–Stokes equation can be expressed as Eq. (A-10).

$$\left( \frac{\partial^2 u}{\partial y^2} + \frac{\partial^2 u}{\partial z^2} \right) = \frac{\partial p}{\mu \cdot \partial x} \tag{A-10}$$

The boundary conditions for the $SiO_2$ nanochannels will be nonslip, whereas the boundary condition for graphene nanochannels is a slip boundary conditions on the left, right, and bottom of the nanochannel, with the nonslip boundary condition on the top side covered by Pyrex glass. First, we must assume that the velocity profile is composed of $u_1$ and $u_2$, as shown in Eq. (A-11), where $u_1$ represents the velocity profile with the nonslip boundary conditions and $u_2$ represents the velocity profile with the slippery condition. $u_2$ is composed of four different boundary conditions, including $u_{2-1}$ representing the slip velocity $u_{slipa}$ on the top side of the channel, $u_{2-2}$ representing the slip velocity $u_{slipb}$ on the right side of the channel, $u_{2-3}$ representing the slip velocity $u_{slipc}$ on the bottom of the channel, and $u_{2-4}$ representing the slip velocity $u_{slipd}$ on the left side of the channel, as shown in Eq. (A-12) and Figure 4d.

$$u = u_1 + u_2 \tag{A-11}$$

$$u_2 = u_{2-1} + u_{2-2} + u_{2-3} + u_{2-4} \tag{A-12}$$

Solving for $u_1$ is simple because this is a purely nonslip boundary conditions on all of the channel walls, with the boundary condition shown in Eqs. (A-13)–(A-16) and by assuming two different Fourier series for the left and right sides of Eq. (A-10). The velocity profile can be obtained, as shown in Eq. (4) in the main text.

$$z = 0 : u = u_{slipc} = 0 \tag{A-13}$$

$$z = 2h : u = u_{slipa} = 0 \tag{A-14}$$

$$y = 0 : u = u_{slipd} = 0 \tag{A-15}$$

$$y = 2w : u = u_{slipb} = 0 \tag{A-16}$$

The rest of the velocity profile should follow Eq. (A-17). In addition, it can be assumed to be a function of $y$ multiplied by a function of $z$, after considering the different boundary conditions listed in Eqs. (A-18)–(A-25). The slip length can be obtained by taking the derivative of the velocity profile at the wall, as shown in Eq. (9) in the main text. $u_{2-1}$, $u_{2-2}$, $u_{2-3}$, and $u_{2-4}$ are given by Eqs. (5)–(8) in the main text.

$$\mu \left( \frac{\partial^2 u_{2-i}}{\partial y^2} + \frac{\partial^2 u_{2-i}}{\partial z^2} \right) = 0, \ i = 1:4 \tag{A-17}$$

$$z = 0, y = 0, y = 2w : u_{2-1} = 0 \tag{A-18}$$

$$z = 2h : u_{2-1} = u_{slipa} \tag{A-19}$$

$$z = 0, z = 2h, y = 0 : u_{2-2} = 0 \tag{A-20}$$

$$y = 2w : u_{2-2} = u_{slipb} \tag{A-21}$$

$$z = 2h, y = 0, y = 2w : u_{2-3} = 0 \tag{A-22}$$

$$z = 0 : u_{2-3} = u_{slipc} \tag{A-23}$$

$$z = 0, z = 2h, y = 2w : u_{2-4} = 0 \tag{A-24}$$

$$y = 0 : u_{2-4} = u_{slipd} \tag{A-25}$$

With all the velocity profiles in Eq. (3) shown in the main text obtained, the relation between the moving location and time for a nanochannel with the wall having slip boundary conditions can be obtained using the same integration procedure mentioned previously, as shown in Eq. (A-26) and Eq. (A-27), where $D_1$, $D_2$, $D_3$, $D_4$, and $D_5$ are shown in Eqs. (A-28)–(A-32).

$$Q = \frac{dV}{dt} = \int_0^{2h}\int_0^{2w} u(y,z)\,dydz = \int_0^{2h}\int_0^{2w} u_1 + u_{2-1} + u_{2-2} + u_{2-3} + u_{2-4}\,dydz = 2w \cdot 2h \cdot \frac{dl}{dt} \quad \text{(A-26)}$$

$$l = \sqrt{(D_1 + D_2 + D_3 + D_4 + D_5) \cdot t} \quad \text{(A-27)}$$

$$D_1 = \frac{8\Delta p h^2}{\mu w \pi^3} \sum_{n=0}^{\infty} \frac{(-1)^n}{(2n+1)^3}\left(2w - \frac{4h}{(2n+1)\pi}\cdot\tanh\frac{(2n+1)\pi w}{2h}\right)\cdot\left[\frac{4h}{(2n+1)\pi}\cdot\sin\frac{(2n+1)\pi}{2}\right] \quad \text{(A-28)}$$

$$D_2 = \frac{4w u_{slipa}}{h\pi^3} \sum_{n=0}^{\infty} \frac{1}{n^3 \cdot \sin\left(\frac{n\pi h}{w}\right)} \cdot [\cosh(n\pi) - 1]^2 \cdot \left[1 - \cos\frac{n\pi h}{w}\right] \quad \text{(A-29)}$$

$$D_3 = \frac{4h u_{slipb}}{w\pi^3} \sum_{n=0}^{\infty} \frac{1}{n^3 \cdot \sinh\left(\frac{n\pi w}{h}\right)} \cdot (1 - \cos n\pi)^2 \cdot \left[\cosh\left(\frac{n\pi w}{h}\right) - 1\right] \quad \text{(A-30)}$$

$$D_4 = \frac{4w u_{slipc}}{h\pi^3} \sum_{n=0}^{\infty} \frac{1}{n^3 \cdot \sin\left(\frac{n\pi h}{w}\right)} \cdot [\cosh(n\pi) - 1]^2 \cdot \left[1 - \cos\frac{n\pi h}{w}\right] \quad \text{(A-31)}$$

$$D_5 = \frac{4h u_{slipd}}{w\pi^3} \sum_{n=0}^{\infty} \frac{1}{n^3 \sinh\left(\frac{n\pi w}{h}\right)} \cdot (1 - \cos n\pi)^2 \cdot \left[\cosh\left(\frac{n\pi w}{h}\right) - 1\right] \quad \text{(A-32)}$$

The pressure term shown in Eq. (A-28) is the capillary pressure. We consider different contact angles on different walls as shown in Eq. (12) in the main text. For a simpler assumption where the contact angle on all walls is the same, the capillary pressure can be expressed as Eq. (A-33). However, we consider the water in the channel to be moving, which makes it reasonable to use the dynamic contact angle instead of the static contact angle. There are several models of dynamic contact angle that can be used [40-43]. Here, we calculate the dynamic contact angle using Eq. (A-34) [44], where $\theta_s$ is the static contact angle, $\theta_m$ is the dynamic contact angle, and $Ca$ is the capillary number, where $Ca = \frac{\mu u}{\sigma}$. For easier expression, we consider all the contact angles on different channel walls to be the same. Therefore, the pressure term can be expressed as shown in Eq. (A-35).

$$\Delta p = \frac{(w+h) \cdot \sigma \cdot \cos\theta}{wh} \tag{A-33}$$

$$\frac{\cos\theta_s - \cos\theta_m}{\cos\theta_s + 1} = \tanh\left(4.96 Ca^{0.702}\right) \tag{A-34}$$

$$\Delta p = \frac{(w+h)\sigma\left[\cos\theta_s - (\cos\theta_s + 1) \cdot \tanh\left(4.96 Ca^{0.702}\right)\right]}{wh} \tag{A-35}$$

For a nanochannel with all the channel walls having a nonslip boundary conditions, $D_2$, $D_3$, $D_4$, and $D_5$ will be 0. Therefore, the water flow movement in the nanochannels can be expressed by Eq. (A-36).

$$l = \sqrt{\left[\frac{8 \cdot (w+h) \cdot \sigma \cdot \left[\cos\theta_s - (\cos\theta_s + 1) \cdot \tanh\left(4.96 Ca^{0.702}\right)\right] \cdot h}{\mu w^2 \pi^3} \cdot \sum_{n=0}^{\infty} \frac{(-1)^n}{(2n+1)^3}\left(2w - \frac{4h}{(2n+1)\pi} \cdot \tanh\frac{(2n+1)\pi w}{2h}\right) \cdot \left[\frac{4h}{(2n+1)\pi} \cdot \sin\frac{(2n+1)\pi}{2}\right]\right] \cdot t} \tag{A-36}$$

Note that a linear relation between $l$ and $t^{0.5}$ in a fully developed laminar flow can be observed, as shown in Eq. (A-36). However, this is unlikely to occur in the real world because the exact region needed to form the fully developed flow is unknown. In such a case, an entrance length ($l_0$) before the flow becomes fully developed is needed. The corresponding equation then becomes Eq. (A-37). The fully developed flow starts when the time becomes 0.

$$l = \sqrt{\left[\frac{8 \cdot (w+h) \cdot \sigma \cdot \left[\cos\theta_s - (\cos\theta_s + 1) \cdot \tanh\left(4.96 Ca^{0.702}\right)\right] \cdot h}{\mu w^2 \pi^3} \cdot \sum_{n=0}^{\infty} \frac{(-1)^n}{(2n+1)^3}\left(2w - \frac{4h}{(2n+1)\pi} \cdot \tanh\frac{(2n+1)\pi w}{2h}\right) \cdot \left[\frac{4h}{(2n+1)\pi} \cdot \sin\frac{(2n+1)\pi}{2}\right]\right] \cdot t} + l_0 \tag{A-37}$$

## CRediT authorship contribution statement

**Kuan-Ting Chen:** Methodology, Investigation, Data curation, Writing – original draft. **Qin-Yi Li:** Conceptualization, Methodology, Validation, Supervision, Writing – review & editing. **Takeshi Omori:** Methodology, Validation, Writing – review & editing. **Yasutaka Yamaguchi:** Methodology, Validation, Writing – review & editing. **Tatsuya Ikuta:** Methodology, Validation. **Koji Takahashi:** Supervision, Resources, Writing – review & editing.

## Declaration of competing interest

The authors declare that they have no known competing financial interests or personal relationships that could have appeared to influence the work reported in this paper.

## ACKNOWLEDGMENTS

This work was partially supported by the JST CREST Grant Number JPMJCR18I1, Japan, and JSPS KAKENHI (Grant Nos. JP18K03929, JP18K03978, JP20H02089, JP20H02090, and JP21K18693). We thank Mr. Kun CHENG for his assistance with the FEM simulations.

## REFERENCES

(1) Sparreboom, W.; van den Berg, A.; Eijkel, J. C. Principles and applications of nanofluidic transport. *Nature nanotechnology* **2009,** *4* (11), 713-720.

(2) Li, Q. Y.; Matsushita, R.; Tomo, Y.; Ikuta, T.; Takahashi, K. Water confined in hydrophobic cup-stacked carbon nanotubes beyond surface-tension dominance. *The journal of physical chemistry letters* **2019,** *10* (13), 3744-3749.

(3) Holt, J. K.; Park, H. G.; Wang, Y.; Stadermann, M.; Artyukhin, A. B.; Grigoropoulos, C. P.; Noy, A.; Bakajin, O. Fast mass transport through sub-2-nanometer carbon nanotubes. *Science* **2006,** *312* (5776), 1034-1037.

(4) Majumder, M.; Chopra, N.; Andrews, R.; Hinds, B. J. Enhanced flow in carbon nanotubes. *Nature* **2005,** *438* (7064), 44-44.

(5) Thomas, J. A.; McGaughey, A. J. Reassessing fast water transport through carbon nanotubes. *Nano Lett* **2008,** *8* (9), 2788-2793.

(6) Falk, K.; Sedlmeier, F.; Joly, L.; Netz, R. R.; Bocquet, L. Molecular origin of fast water transport in carbon nanotube membranes: superlubricity versus curvature dependent friction. *Nano Lett* **2010,** *10* (10), 4067-4073.

(7) Falk, K.; Sedlmeier, F.; Joly, L.; Netz, R. R.; Bocquet, L. Ultralow liquid/solid friction in carbon nanotubes: Comprehensive theory for alcohols, alkanes, OMCTS, and water. *Langmuir* **2012,** *28* (40), 14261-14272.

(8) Guo, L.; Chen, S.; Robbins, M. O. Slip boundary conditions over curved surfaces. *Phys Rev E* **2016,** *93* (1), 013105.

(9) Joly, L. Capillary filling with giant liquid/solid slip: dynamics of water uptake by carbon nanotubes. *The Journal of chemical physics* **2011,** *135* (21), 214705.

(10) Thomas, J.; McGaughey, A. Density, distribution, and orientation of water molecules inside and outside carbon nanotubes. *The Journal of chemical physics* **2008,** *128* (8), 084715.

(11) Stokes, G. G. On the theories of the internal friction of fluids in motion, and of the equilibrium and motion of elastic solids. *Transactions of the Cambridge Philosophical Society* **1880,** *8*.

(12) Kumar Kannam, S.; Todd, B.; Hansen, J. S.; Daivis, P. J. Slip length of water on graphene: Limitations of non-equilibrium molecular dynamics simulations. *The Journal of chemical physics* **2012,** *136* (2), 024705.

(13) Wagemann, E.; Oyarzua, E.; Walther, J. H.; Zambrano, H. A. Slip divergence of water flow in graphene nanochannels: the role of chirality. *Phys Chem Chem Phys* **2017,** *19* (13), 8646-8652.

(14) Wagemann, E.; Becerra, D.; Walther, J. H.; Zambrano, H. A. Water flow enhancement in amorphous silica nanochannels coated with monolayer graphene. *MRS Communications* **2020,** *10* (3), 428-433.

(15) Xie, Q.; Alibakhshi, M. A.; Jiao, S.; Xu, Z.; Hempel, M.; Kong, J.; Park, H. G.; Duan, C. Fast water transport in graphene nanofluidic channels. *Nature nanotechnology* **2018,** *13* (3), 238-245.

(16) Washburn, E. W. The dynamics of capillary flow. *Physical review* **1921,** *17* (3), 273.

(17) Lucas, R. Ueber das Zeitgesetz des kapillaren Aufstiegs von Flüssigkeiten. *Kolloid-Zeitschrift* **1918,** *23* (1), 15-22.

(18) Keerthi, A.; Goutham, S.; You, Y.; Iamprasertkun, P.; Dryfe, R. A.; Geim, A. K.; Radha, B. Water friction in nanofluidic channels made from two-dimensional crystals. *Nat Commun* **2021,** *12* (1), 1-8.

(19) Chang, C. C.; Yang, R. J.; Wang, M.; Miau, J. J.; Lebiga, V. Liquid flow retardation in nanospaces due to electroviscosity: Electrical double layer overlap, hydrodynamic slippage, and ambient atmospheric $CO_2$ dissolution. *Phys Fluids* **2012,** *24* (7), 072001.

(20) Haneveld, J.; Tas, N. R.; Brunets, N.; Jansen, H. V.; Elwenspoek, M. Capillary filling of sub-10 nm nanochannels. *J Appl Phys* **2008,** *104* (1), 014309.

(21) Phan, V. N.; Yang, C.; Nguyen, N. T. Analysis of capillary filling in nanochannels with electroviscous effects. *Microfluid Nanofluid* **2009,** *7* (4), 519.

(22) Tas, N. R.; Haneveld, J.; Jansen, H. V.; Elwenspoek, M.; van den Berg, A. Capillary filling speed of water in nanochannels. *Appl Phys Lett* **2004,** *85* (15), 3274-3276.

(23) Tas, N. R.; Mela, P.; Kramer, T.; Berenschot, J. W.; van den Berg, A. Capillarity induced negative pressure of water plugs in nanochannels. *Nano Lett* **2003,** *3* (11), 1537-1540.

(24) Van Honschoten, J.; Escalante, M.; Tas, N. R.; Elwenspoek, M. Formation of liquid menisci in flexible nanochannels. *J Colloid Interf Sci* **2009,** *329* (1), 133-139.

(25) Yang, M.; Cao, B. Y.; Wang, W.; Yun, H. M.; Chen, B. M. Experimental study on capillary filling in nanochannels. *Chemical Physics Letters* **2016,** *662*, 137-140.

(26) Cao, B. Y.; Yang, M.; Hu, G.-J. Capillary filling dynamics of polymer melts in nanopores: experiments and rheological modelling. *Rsc Adv* **2016,** *6* (9), 7553-7559.

(27) Suk, J. W.; Kitt, A.; Magnuson, C. W.; Hao, Y.; Ahmed, S.; An, J.; Swan, A. K.; Goldberg, B. B.; Ruoff, R. S. Transfer of CVD-grown monolayer graphene onto arbitrary substrates. *Acs Nano* **2011,** *5* (9), 6916-6924.

(28) Li, Q. Y.; Xia, K.; Zhang, J.; Zhang, Y.; Li, Q.; Takahashi, K.; Zhang, X. Measurement of specific heat and thermal conductivity of supported and suspended graphene by a comprehensive Raman optothermal method. *Nanoscale* **2017,** *9* (30), 10784-10793.

(29) Li, Q. Y.; Takahashi, K.; Ago, H.; Zhang, X.; Ikuta, T.; Nishiyama, T.; Kawahara, K. Temperature dependent thermal conductivity of a suspended submicron graphene ribbon. *J Appl Phys* **2015,** *117* (6), 065102.

(30) Kozbial, A.; Li, Z.; Conaway, C.; McGinley, R.; Dhingra, S.; Vahdat, V.; Zhou, F.; D'Urso, B.; Liu, H.; Li, L. Study on the surface energy of graphene by contact angle measurements. *Langmuir* **2014,** *30* (28), 8598-8606.

(31) Schwiebert, M. K.; Leong, W. H. Underfill flow as viscous flow between parallel plates driven by capillary action. *IEEE Transactions on Components, Packaging, and Manufacturing Technology: Part C* **1996,** *19* (2), 133-137.

(32) Stange, M.; Dreyer, M. E.; Rath, H. J. Capillary driven flow in circular cylindrical tubes. *Phys Fluids* **2003,** *15* (9), 2587-2601.

(33) Jennings, S. The mean free path in air. *Journal of Aerosol Science* **1988,** *19* (2), 159-166.

(34) Yang, D.; Krasowska, M.; Priest, C.; Popescu, M. N.; Ralston, J. Dynamics of capillary-driven flow in open microchannels. *The Journal of Physical Chemistry C* **2011,** *115* (38), 18761-18769.

(35) Tocci, G.; Joly, L.; Michaelides, A. Friction of water on graphene and hexagonal boron nitride from ab initio methods: very different slippage despite very similar interface structures. *Nano Lett* **2014,** *14* (12), 6872-6877.

(36) Joseph, S.; Aluru, N. Why are carbon nanotubes fast transporters of water? *Nano Lett* **2008,** *8* (2), 452-458.

(37) Greenwood, G.; Kim, J. M.; Zheng, Q.; Nahid, S. M.; Nam, S.; Espinosa-Marzal, R. M. Effects of Layering and Supporting Substrate on Liquid Slip at the Single-Layer Graphene Interface. *Acs Nano* **2021**.

(38) Celebi, A. T.; Barisik, M.; Beskok, A. Surface charge-dependent transport of water in graphene nano-channels. *Microfluid Nanofluid* **2018,** *22* (1), 1-10.

(39) Thamdrup, L. H.; Persson, F.; Bruus, H.; Kristensen, A.; Flyvbjerg, H. Experimental investigation of bubble formation during capillary filling of Si O 2 nanoslits. *Appl Phys Lett* **2007,** *91* (16), 163505.

(40) Blake, T.; Haynes, J. Kinetics of liquidliquid displacement. *J Colloid Interf Sci* **1969,** *30* (3), 421-423.

(41) Cox, R. The dynamics of the spreading of liquids on a solid surface. Part 1. Viscous flow. *J Fluid Mech* **1986,** *168*, 169-194.

(42) Bracke, M.; De Voeght, F.; Joos, P. The kinetics of wetting: the dynamic contact angle. *Trends in Colloid and Interface Science III* **1989**, 142-149.

(43) Popescu, M. N.; Ralston, J.; Sedev, R. Capillary rise with velocity-dependent dynamic contact angle. *Langmuir* **2008,** *24* (21), 12710-12716.

(44) Jiang, T. S.; Soo Gun, O.; Slattery, J. C. Correlation for dynamic contact angle. *J Colloid Interf Sci* **1979,** *69* (1), 74-77.

# SUPPLEMENTARY INFORMATION

# Slip Length Measurement in Rectangular Graphene Nanochannels with a 3D Flow Analysis

Kuan-Ting Chen[1], Qin-Yi Li[1,4,*], Takeshi Omori[2], Yasutaka Yamaguchi[3,5], Tatsuya Ikuta[1,4], and Koji Takahashi[1,4,*]

[1] Department of Aeronautics and Astronautics, Kyushu University, Fukuoka 819-0395, Japan

[2] Department of Mechanical Engineering, Osaka City University, 3-3-138 Sugimoto, Sumiyoshi, Osaka, Osaka 558-8585, Japan

[3] Department of Mechanical Engineering, Osaka University, 2-1 Yamadaoka, Suita 565-0871, Japan

[4] International Institute for Carbon-Neutral Energy Research (WPI-I2CNER), Kyushu University, Japan

[5] Water Frontier Research Center (WaTUS), Research Institute for Science & Technology, Tokyo University of Science, 1-3 Kagurazaka, Shinjuku-ku, Tokyo 162-8601, Japan

[*]Corresponding authors, Qin-Yi Li: qinyi.li@aero.kyushu-u.ac.jp

Koji Takahashi: takahashi@aero.kyushu-u.ac.jp

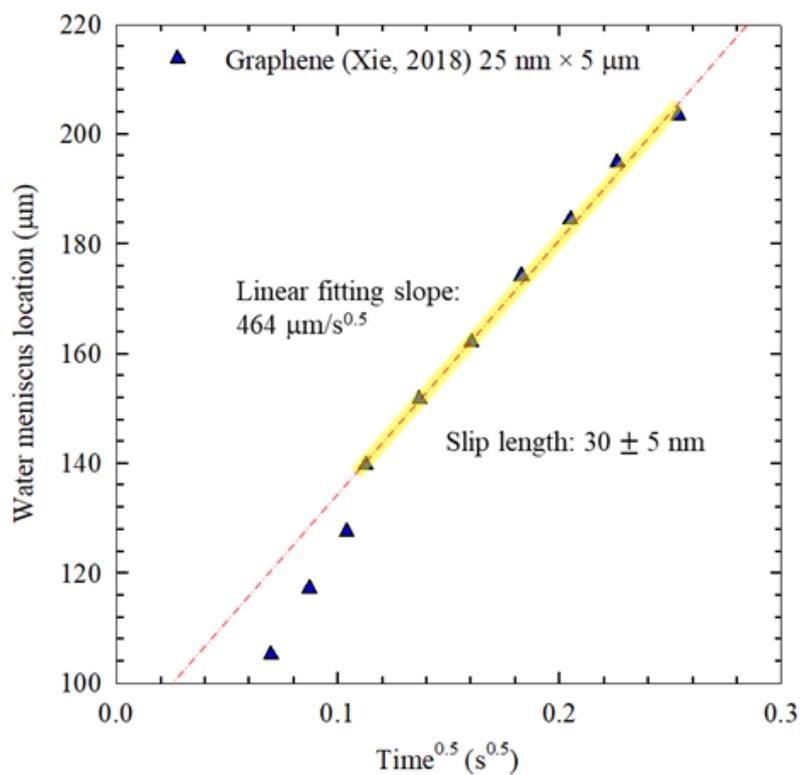

**Figure S6** Fitting of the meniscus location in the fully developed flow region in a 25-nm-deep graphene nanochannel using the data reported by Xie *et al.* [1].

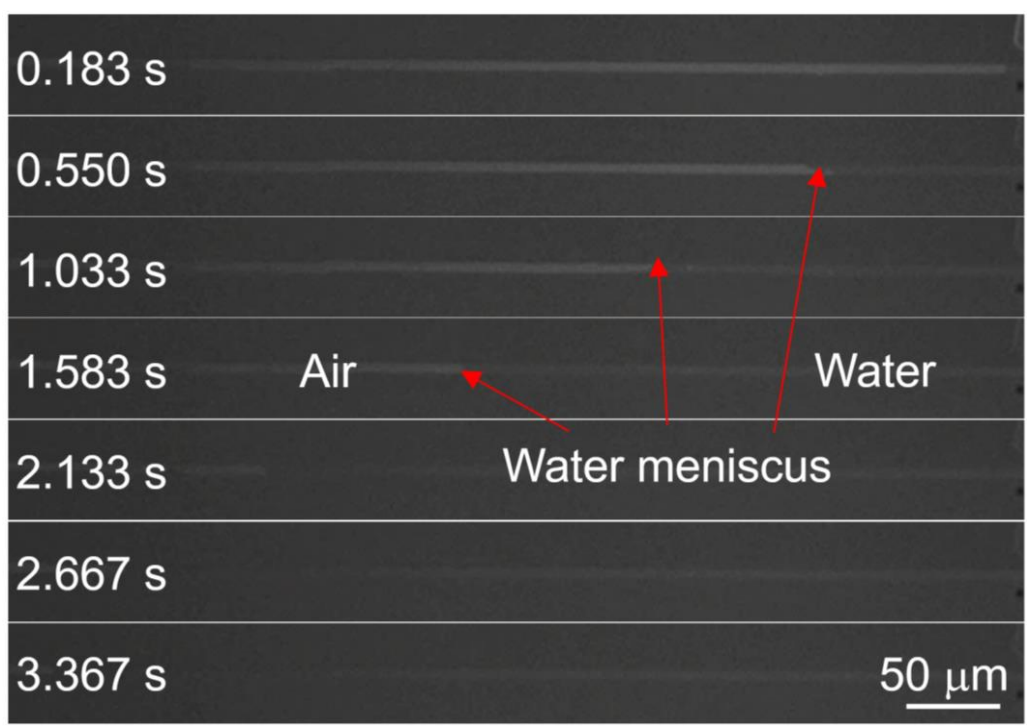

**Figure S7** High-speed-camera images of capillary filling in the second 45-nm-deep and 5-µm-wide graphene nanochannel (Gr channel2).

**Supplementary Note S1 Verification of the analytical solution to the 3D flow model**

The analytical solution to the 3D flow model was verified by comparing the analytical results with finite element method (FEM) simulations using the commercial software of COMSOL. The studied channel for the simulations is 45 nm in depth, 50 nm in width, and 50 nm in length. The simulation domain has a total element number of 2928189 as shown in Figure S8. The meshes are denser at the boundaries to calculate the boundary area accurately. Water was given an initial velocity of 0.1 μm/s at the channel entrance. The pressure gradient along the channel was $1.2 \times 10^{11}$ Pa/m, which is the same as the analytical calculation. The difference between the analytical results and the FEM simulation results of the vertical velocity profile at the middle of the channel is 0.4 to 0.6%, which shows the accuracy of our model. The final results of the velocity profile for two surfaces in the channel were shown in Figure S10. We also compare the analytical results with 2D FEM simulation in an extreme condition with a channel with a width/depth ratio of 10000 as shown in Figure S11. The two-dimensional velocity profile in the *x-z* surface was done by FEM simulation as shown in Figure S11a, and the velocity profile matches well with the 3D model in extreme channel conditions as shown in Figure S11b.

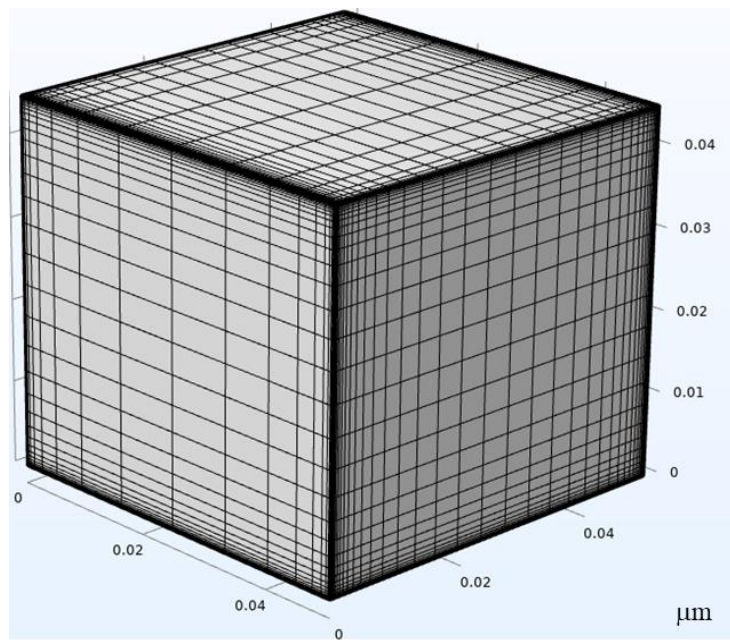

**Figure S8** Mesh for the channel with a height of 45 nm and width of 50 nm.

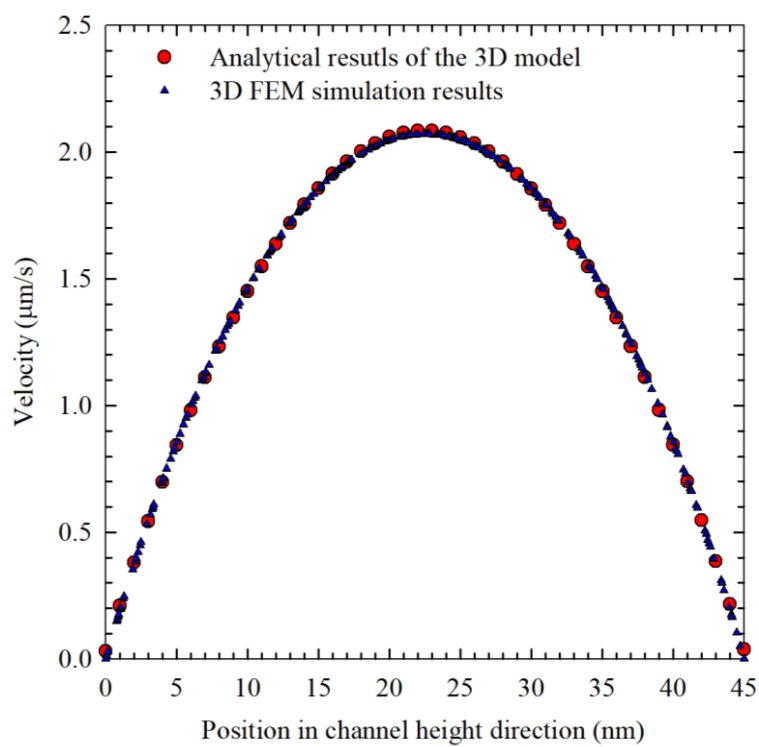

**Figure S9** Comparison of analytical results and simulation results on vertical velocity profile in a channel with a height of 45 nm and width of 50 nm.

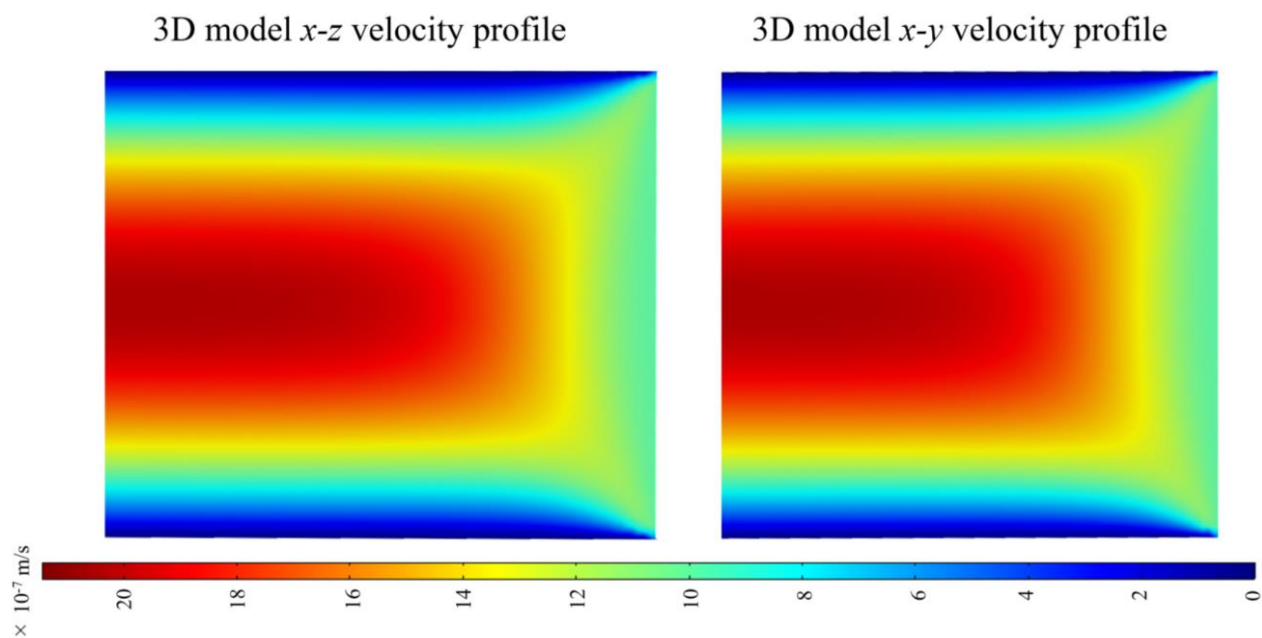

**Figure S10** Simulation results of 3D model velocity profile in the *x-z* surface and the *x-y* surface in a ch-annel with a height of 45 nm and width of 50 nm.

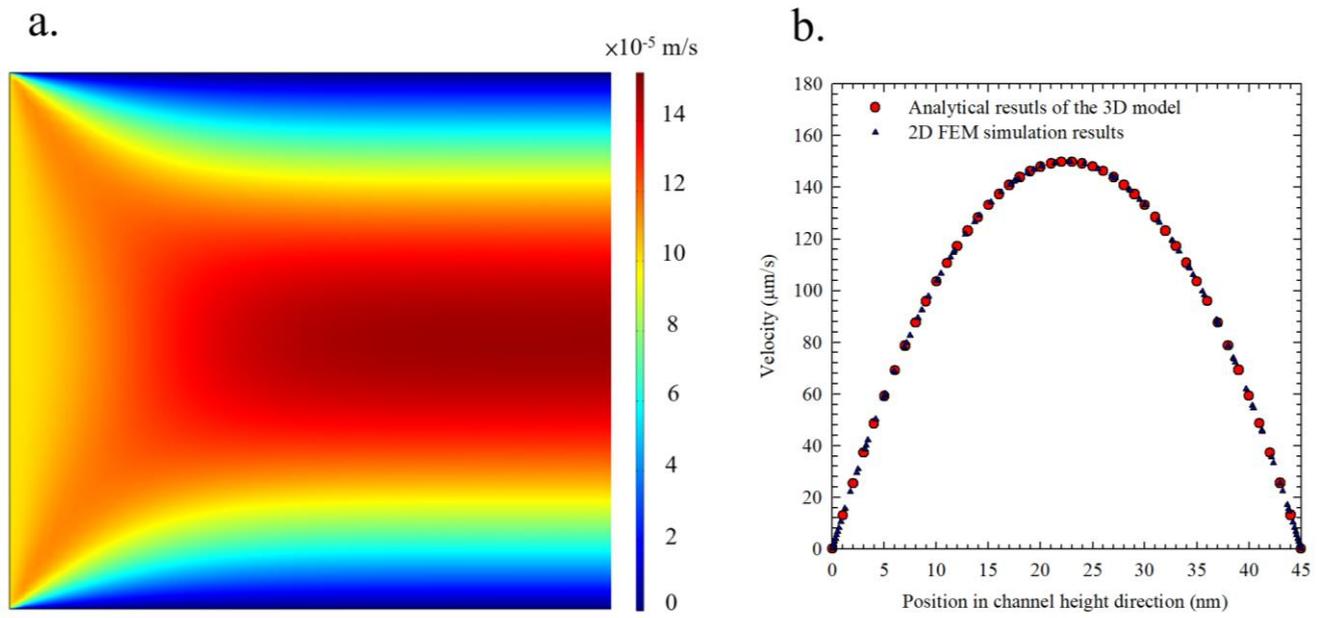

**Figure S11** Simulation results in a 2D channel. (a) Velocity profile in the *x-z* surface in a 2D channel wi-th a height of 45 nm. (b) Velocity profile in a *x-z* surface for a channel with a width/depth ratio of 1000-0 calculated by the 3D model compared with the FEM simulation.

**Supplementary Note S2 Uncertainty Analysis**

From Eq. (9) in the main text, we know that the slip length was determined from the slip velocity and slope of the velocity profile at the wall. Denoting the relative uncertainty of the slip velocity and the slope of the velocity profile on the wall as $E_{uslip}$ and $E_{du}$, respectively, we can express the relative uncertainty of the slip length ($E_{sliplength}$) as Eq. (S1).

$$E_{sliplength} = \sqrt{\left(E_{uslip}\right)^2 + \left(E_{du}\right)^2} \tag{S1}$$

The slip velocity can be determined by observing the meniscus movement during capillary filling as shown in Eq. (A-27)–(A-32) in the Appendix. The top side of the channel was covered by Pyrex glass without slip condition, which means $D_2$ is 0, and the slip velocities on the other three walls were the same ($u_{slip}$). For easy expression, we write $D_3 = u_{slip} \cdot A$, $D_4 = u_{slip} \cdot B$, $D_5 = u_{slip} \cdot C$. The slip velocity can be expressed as Eq. (S2) and the corresponding uncertainty can be expressed as Eq. (S3).

$$u_{slip} = \frac{\frac{l^2}{t} - D_1}{A+B+C} \tag{S2}$$

$$E_{u_{slip}} = \sqrt{\left(E_{\frac{l^2}{t}} + E_{D_1}\right)^2 + \left(E_A + E_B + E_C\right)^2} \tag{S3}$$

The uncertainty of $\frac{l^2}{t}$ represents the uncertainty of the linear fitting; the uncertainties of $D_1$, $A$, $B$, $C$ come from $\Delta p$, $h$, $w$, $\mu$. Because the resolution of the AFM measurement was 0.5 nm, the relative uncertainty of the channel depth ($E_h$) could be obtained by Eq. (S4); the channel depth in this work was 45 nm and the corresponding uncertainty was 1.1%.

$$E_h = \sqrt{\left(\frac{\Delta h}{h}\right)^2} \tag{S4}$$

The uncertainty of $\Delta p$ ($E_{\Delta p}$) can be determined from Eq. (12) in the main text, which is expressed here as Eq. (S5).

$$\Delta p = \frac{\sigma(\cos\theta_1 + \cos\theta_2)}{2h} + \frac{\sigma(\cos\theta_3 + \cos\theta_4)}{2w} \tag{S5}$$

The uncertainties of surface tension, channel width, and contact angle were ignored. Thus, the remaining uncertainty comes from the channel depth. Because there is only one term of channel depth in Eq. (S5), the uncertainty of $\Delta p$ is 1.1%.

Having determined $E_{\Delta p}$ and $E_h$, $E_{D_1}$ can be calculated by Eq. (S6), which is derived from Eq. (A-28) in the Appendix:

$$E_{D_1} = \sqrt{(E_{\Delta p})^2 + (2 \cdot E_h)^2} = 1.91\% \tag{S6}$$

We neglected the uncertainty inside the series because it was calculated with more than $10^{20}$ terms. $E_A$, $E_B$, and $E_C$ only have one term of the channel depth term outside the series as can be seen in Eq. (A-30) – (A-32) in the Appendix. This leads to the aforementioned uncertainty of 1.1%.

The uncertainty of the linear fitting slope ($E_{\frac{l^2}{t}}$) was acquired after linear fitting as 1.2%. With $E_{\frac{l^2}{t}}$, $E_{D_1}$, $E_A$, $E_B$, and $E_C$ obtained, the uncertainty in the slip velocity ($E_{u_{slip}}$) was calculated as 4.5% using Eq. (S2). The uncertainty of the slope of the velocity profile ($E_{du}$) on the wall can be calculated with the same technique based on the derivative of Eqs. (3) – (8) in the main text. Because $u_{slipa}$ is zero, $u_{2-1}$ will be zero. We assume the derivative of each velocity to be $\frac{\partial u_{2-2}}{\partial z} = u_{slip} \cdot E$, $\frac{\partial u_{2-3}}{\partial z} = u_{slip} \cdot F$, $\frac{\partial u_{2-4}}{\partial z} = u_{slip} \cdot G$. The derivative of the velocity profile can be expressed as Eq. (S7) and the uncertainty of the slope of the velocity profile on the wall ($E_{du}$) can be expressed as Eq. (S8).

$$\frac{\partial u}{\partial z} = \frac{\partial u_1}{\partial z} + u_{slip} \cdot (E + F + G) \tag{S7}$$

$$E_{du} = \sqrt{\left[ E_{\frac{\partial u_1}{\partial z}}^2 + \left( E_E + E_F + E_G \right)^2 + E_{u_{slip}}^2 \right]} \qquad (S8)$$

where $E_{\frac{\partial u_1}{\partial z}}$, $E_E$, $E_F$, and $E_G$ are functions of $\Delta p$ and $h$, and the $Eu_{slip}$ was obtained from Eq. (S4). With all of the uncertainties in Eq. (S8) acquired, $E_{du}$ was then calculated as 8.3%. Finally, the total uncertainty for the slip length was calculated as 9.4% with Eq. (S1). The uncertainty of the corrected slip length using the experimental data of Xie et al. [1] was calculated using the same technique assuming the same uncertainty for AFM measurements because the author did not state this in their paper. The results are shown in Table S1. Because Xie et al. [1] only gave the uncertainty of the resistance ratio in their research (4.3%), further analysis is needed to acquire the uncertainty of the slip length. In Eq. (3) in their paper [1], the slip length is a function of resistance ratio, channel width, channel depth, viscosity, and density. By considering the uncertainty of the channel depth (2%) and resistance ratio (4.3%) while neglecting other parameters, an uncertainty of 4.6% can be calculated. In the research of Keerthi et al. [2], the channel depth was determined by the number of graphene layers. This means that the only parameter that can affect the uncertainty of the slip length is the weight loss rate, which had an uncertainty of 34.0% in their work [2].

Table S1 Uncertainties in the results of previous studies.

|  | Gr channel1 | Gr channel2 | Xie et al.[16] | Xie et al.[1] (corrected) | Keerthi et al.[2] | Keerthi et al.[2] (corrected) |
|---|---|---|---|---|---|---|
| **Total uncertainties** | 9.4% | 9.5% | 4.6% | 16.7% | 34.0% | 34.0% |
| $E_{du}$ | 8.3% | 8.3% |  | 14.3% |  |  |
| $E_{uslip}$ | 4.5% | 4.5% |  | 8.6% |  |  |
| $E_h$ | 1.1% | 1.1% | 2% | 2% |  |  |
| $E_{\Delta p}$ | 1.1% | 1.1% |  | 2% |  |  |
| $E_{Slope}$ | 1.2% | 1.2% |  | 2.7% |  |  |

# REFERENCES

(1) Xie, Q.; Alibakhshi, M. A.; Jiao, S.; Xu, Z.; Hempel, M.; Kong, J.; Park, H. G.; Duan, C. Fast water transport in graphene nanofluidic channels. *Nature nanotechnology* **2018,** *13* (3), 238-245.

(2) Keerthi, A.; Goutham, S.; You, Y.; Iamprasertkun, P.; Dryfe, R. A.; Geim, A. K.; Radha, B. Water friction in nanofluidic channels made from two-dimensional crystals. *Nat Commun* **2021,** *12* (1), 1-8.